\documentstyle[floats,multicol,aps,epsf,prb]{revtex}

\newcommand \bea {\begin{eqnarray} }
\newcommand \eea {\end{eqnarray}}

\newcommand \dg {^{\dagger}}
\begin{document}
\draft
\twocolumn[\hsize\textwidth\columnwidth\hsize\csname @twocolumnfalse\endcsname
\title{
Marginal Fermi Liquid in a lattice of three-body bound-states. 
}
\author{ A. F. Ho$^{1,2}$ and  P. Coleman$^{1,2}$ }
\address{
$^1$Serin Laboratory, Rutgers University, P.O. Box 849,
Piscataway, New Jersey 08855-0849}
\address{
$^2$Department of  Physics, University of Oxford, 1 Keble Road, 
Oxford OX1 3NP,  UK}
\maketitle
\date{\today}
\maketitle
\begin{abstract}
We study a lattice model for Marginal Fermi liquid behavior, involving
a gas of electrons coupled to a dense lattice of three-body
bound-states.  The presence of the bound-states changes the phase
space for electron-electron scattering and induces a marginal
self-energy amongst the electron gas.  When the three-body
bound-states are weakly coupled to the electron gas, there is a
substantial window for marginal Fermi liquid behavior and in this
regime, the model displays the presence of two relaxation times, one
linear, one quadratic in the temperature.  At low temperatures the
bound-states develop coherence leading to a cross-over to conventional
Fermi liquid behavior.  At strong-coupling, marginal Fermi liquid
behavior is pre-empted by a pairing or magnetic instability, and it is
not possible to produce a linear scattering rate
comparable with the temperature.  We discuss the low temperature
instabilities of this model and compare it to the Hubbard model at
half-filling.
\end{abstract}

\vskip 0.2 truein
\vskip2pc]

\section{INTRODUCTION}

The concept of a marginal Fermi liquid (MFL)  
was invented as a phenomenological
description of the asymptotic
properties of 
of high temperature superconductors in their normal state\cite{Varma}. These include
a linear 
electrical resistivity $\rho \sim T $ over at least 2 decades in
temperature and a  quasi-particle scattering rate which may be
proportional to frequency up to energies as high as 750 meV.\cite{ibm,french_group}
Even in underdoped cuprate superconductors, 
optical data indicates that marginal
behavior develops at scales above the spin gap.\cite{timusk}
We shall take a marginal Fermi liquid to be a system of fermions with
an inelastic scattering rate of the form
\bea
\Gamma\propto {\rm max}[\mid \omega \mid , T],
\eea
where there is no significant momentum dependence of $\Gamma$. In the
cuprates, the constant of proportionality is of order unity.
Analyticity then ensures that the appropriate self-energy 
takes the form
\bea
 \Sigma (\omega) \sim \omega \ln {\rm max}[\mid \omega \mid, T] .
\eea
Many proposals have been made to explain the origin of this
unusual behavior. One prevalent idea, is that  the marginal
Fermi liquid behavior derives from scattering off a soft bosonic mode. 
This idea underpins   the Van Hove scenario\cite{VanHove}, 
 gauge theory models\cite{LeeIoffe} and the quantum-critical 
scattering\cite{Pines} picture
of the cuprates. By contrast, Anderson\cite{Lutt} proposes that the
cuprate metal is a  fully developed Luttinger liquid, with 
 power-law self-energies
which have been mis-identified as a logarithm. Since the soft-mode theories
furnish an electron self-energy which is strongly momentum dependent,
none of these proposals actually gives rise to a marginal
Fermi liquid as originally envisaged.

In this paper, we return to the original proposal,
asking whether a marginal Fermi liquid can form
in a dense electronic system. We pursue an  early
speculation, due to Ruckenstein and Varma \cite{RuckVarma}, that the
marginal self-energy might derive from the scattering of conduction
 electrons off a dispersionless localized bound-state $\Phi$
at the Fermi energy, giving rise to the interaction: 
\bea
H_{int} = \lambda 
(c^{+}_{i \uparrow} c_{i \uparrow} c^{+}_{i \downarrow} \Phi + h.c. )  .
\eea
The presence of such states precisely at the Fermi surface would mean that  
the three-particle phase space grows linearly
with energy.   Inelastic scattering into the localized bound-state
then leads to a marginal self-energy in leading order perturbation
theory, as illustrated in Fig. 1. (The localised state is represented
by the dash line.)  Since the hypothetical
object at the Fermi surface scatters electrons in triplets,
Ruckenstein and Varma identified it as a three-body bound state.


\begin{figure}[here]
\unitlength1.0cm
\begin{center}
\begin{picture}(7,3)
\epsfxsize=7.0cm
\epsfysize=3.0cm
\epsfbox{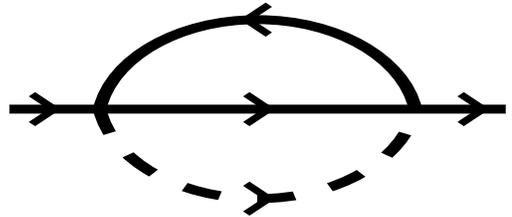}
\end{picture}
\vskip 0.3 truein
\caption{Marginal self-energy diagram for the band electrons. 
Notation: thick lines denote band electron propagators, dashed
lines denote the localised state propagator.}
\end{center}
\end{figure}

\begin{figure}[here]
\unitlength1.0cm
\begin{center}
\begin{picture}(7,3)
\epsfxsize=7.0cm
\epsfysize=3.0cm
\epsfbox{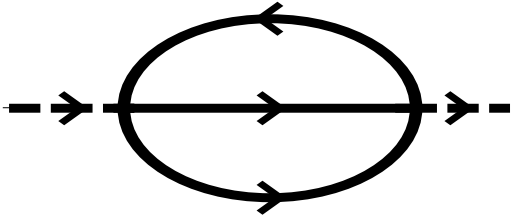}
\end{picture}
\vskip 0.3 truein
\caption{Self-energy for the localised state.}
\end{center}
\end{figure}

A great difficulty with this picture is that it cannot 
be made self-consistent. At the same level of perturbation theory
that furnishes a marginal self-energy, the three-body state must
scatter off the electrons to produce a self-energy term of the
form shown in Fig 2.   This self-energy correction inevitably
moves the resonance of the three-body bound-state away from the 
Fermi energy, introducing an unwanted scale into the problem
and causing the singular scattering to disappear.

An unexpected resolution of this problem recently appeared in the
context of the single impurity two-channel Kondo model\cite{2CCK}.
Marginal Fermi liquid behavior does develop in this model, and the
mechanism by which it occurs is remarkably close to the original
three-body bound-state proposal, with one critical
difference: in the two-channel Kondo model, the three-body
bound-state formed at the impurity site carries no internal quantum
numbers (spin and charge). The associated bound-state fermion is 
represented by a Hermitian operator:
\bea
\Phi = \Phi^{\dagger} .
\eea
This type of fermion is known as a  Majorana fermion\cite{Majorana} 
\footnote{Such an 
object can always be constructed as a linear combination of
two charged fermions $\Phi = \frac{1}{\sqrt{2}}(a + a\dg)$.}.
The effective action for such a field must have the form:
\bea
S = \int d\omega \Phi(-\omega)
(\omega - \Sigma (\omega)) \Phi(\omega)  . 
\eea
Since $\Phi(-\omega) \Phi(\omega) = - \Phi(\omega) \Phi(-\omega)$ due to 
the Grassmanian nature of $\Phi$,
$\Sigma (\omega)$ will be an odd function of frequency, so that
$\Sigma (0) = 0$. 
In other words, the particle-hole symmetry of the $\Phi$ field guarantees
that its energy is pinned to the
Fermi surface. 

In the two-channel Kondo model (after dropping the charge degrees of freedom,
thanks to spin-charge decoupling\cite{Emery}), the total spin  $\vec{S}$
of the two conduction channels at the impurity site can be written in the form:
$\vec{S} = -\frac{i}{2} \vec{\Psi}(0) \times \vec{\Psi}(0)$ where $\vec{\Psi} =
(\Psi^{(1)}, \Psi^{(2)}, \Psi^{(3)})$ is a triplet of Majorana fermions. These
three high energy degrees of freedom bind at the impurity site to form the localised
three-body bound-state $\Phi (0)$, represented as the contraction of the three fermions:
\bea
 \Psi^{(1)}(0) \Psi^{(2)}(0) \Psi^{(3)}(0)  = A \Phi (0) ,
\eea where $A$ is the amplitude for forming this pole $\Phi$.
The residual interaction with the bulk spin degrees of freedom in the
low energy world then gives rise to a vertex of the form $\lambda \Psi^{(1)}
\Psi^{(2)} \Psi^{(3)} \Phi$. The challenge here is to see if such a mechanism
could be generalised to a more realistic lattice model. The work described 
below is an attempt to make a first step in this 
direction.

\section{Construction of Model}

We now use these ideas to motivate a simple {\em lattice} model of a marginal Fermi
Liquid. First note that  the Hubbard model at half-filling can be rewritten in a 
Majorana fermion representation\cite{Affleck}, by the following two steps. The Hubbard model is:
\bea
H_{Hubbard} & = & t \sum_{i,a, \sigma} (c^{\dagger}_{i+a ,\sigma}
c_{i, \sigma} + H.c.) \nonumber \\
 & + & U \sum_{i} ( c^{\dagger}_{i \uparrow} c_{i \uparrow} - 1/2 ) 
( c^{\dagger}_{i \downarrow} c_{i \downarrow} - 1/2 )  .
\eea In the first step, we assume that the lattice has a bipartite structure, and
do a gauge transformation on electron operators belonging to one
sublattice of the bipartite lattice: $c_{i\sigma} \rightarrow -i c_{i\sigma}$. Also
let the hopping term connect a sublattice A site only to a sublattice B site, 
and not to other sublattice A sites.
Then the  interaction is unchanged, but the kinetic energy becomes:
\bea
H_{K.E.} = i t \sum_{i,a, \sigma} (c^{\dagger}_{i+a ,\sigma}
c_{i, \sigma} + H.c.) .
\eea Next, rewrite the electron operators using
$c_{\uparrow}=\frac{1}{\surd 2} (\Psi^{(1)} - i \Psi^{(2)}),
c_{i\downarrow}= - \frac{1}{\surd 2} (\Psi^{(3)} + i \Psi^{(0)})$
where the $\Psi^{(a)}$'s are Majorana fermions,  and we get: 
\begin{eqnarray}
H_{Hubbard} & = & i t \sum_{i,c} \sum_{a=0}^3  \Psi^{(a)}_{i+c}
\Psi^{(a)}_{i} \nonumber \\ 
 & & -U \sum_i \Psi^{(0)}_i \Psi^{(1)}_i \Psi^{(2)}_i \Psi^{(3)}_i .
\end{eqnarray}(Note the sign change for the interaction.) 
In Majorana representation, we see explicitly the $SO(4)$
symmetry\cite{Affleck}\cite{Yang} of the Hubbard model at half-filling.

The crucial generalisation of this paper is to break
the $SO(4)$ symmetry down to $O(3)$, setting:
\bea
t \rightarrow t_a  =  \left\{ \begin{array}{r@{\quad:\quad}l}
		t & a=1,2,3 \\
		t_0 & a=0 .  \end{array} \right. .
\eea 
With $\Phi_i \equiv \Psi^{(0)}_i$,
\bea
H  & = & i t_0 \sum_{i,c} \Phi_{i+c} \Phi_i
+ i t \sum_{i,c} \sum_{a=1}^3  \Psi^{(a)}_{i+c}
\Psi^{(a)}_{i} \nonumber \\ 
 & & -U \sum_i \Phi^{(0)}_i \Psi^{(1)}_i \Psi^{(2)}_i \Psi^{(3)}_i .
\eea  When $t_0 =0$, this Hamiltonian describes a lattice of 
localised three-body bound-states $\Phi_i$ coupled to the continuum.

This toy model provides a simple system
to study the properties of these pre-formed bound-states at each site.
A microscopic model would
provide an explanation of the origin of this symmetry breaking
field leading to formation of these bound-states. In this work, we do 
not address this issue, but instead,
ask whether the single impurity marginal Fermi Liquid mechanism 
survives in the lattice, in this model of reduced symmetry.

We shall show that the main physical effect of the lattice (in the
absence of a broken symmetry phase) is that the
previously localised mode $\Phi_i$ can now move from site to site via
virtual fluctuations into the Fermi Sea: see Fig.3,
thereby providing the lattice coherence energy scale $t_0
\sim U^2/t $, below which this marginal Fermi Liquid  reverts to  a
Fermi Liquid. Unfortunately, this means that the marginal Fermi
liquid phenomenology
does not persist to large $U$, which makes the application of our model
to the cuprates rather problematic.

\begin{figure}[here]
\unitlength1.0cm
\begin{center}
\begin{picture}(7,2.5)
\epsfxsize=7.0cm
\epsfbox{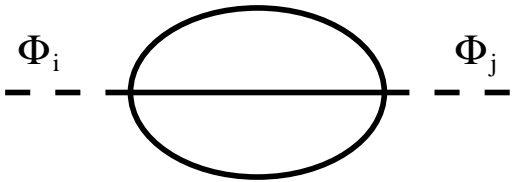}
\end{picture}
\vskip 0.3 truein
\caption{Leading order diagram that generates a dispersion for the $\Phi_i$
fermion. Note that there are no arrows on the propagator lines, as all fermions
are represented by Hermitian operators. }
\end{center}
\end{figure}

To probe the marginal fermi liquid phenomenology, we have calculated the
``optical conductivity''. This cannot be the ordinary electrical current,
because the Hamiltonian of Eqn.15 does not conserve particle
number unless $t_0 = t$ (this is most easily seen with original electron
operators $c, c^+$). The model does have
the $O(3)$ symmetry, so there is the conserved quantity $S^a = -\frac{1}{2}
\sum_j \sum_{\alpha, \beta=1}^{3} 
\Psi^{(\alpha)}_j T^a_{\alpha \beta} \Psi^{(\beta)}_j $, 
where $T^a$ are the three generators of $O(3)$. In 
the representation where $T^a_{\alpha \beta} = i \epsilon_{a \alpha \beta}$,
\bea
S^a = -\frac{i}{2} \sum_j \epsilon_{a b c} \Psi^{(b)}_j \Psi^{(c)}_j .
\eea This leads to the conserved (Noether) current:
\bea
j^a_{i+{\hat x}} = it \epsilon_{a b c} \Psi^{(b)}_{i+{\hat x}} \Psi^{(c)}_i .
\eea (See Methods for its derivation.) One can then define a ``conductivity''
which is the linear response of this $O(3)$ current to an applied field. We 
shall show that this $O(3)$ conductivity has the classic marginal Fermi liquid
behaviour.

Within the marginal Fermi liquid regime, the $\Psi^{(a)}$ retains its 
marginal self-energy. The $\Phi$ mode, while having no self-energy corrections
at $T=0$ and $\omega = 0$, acquires a Fermi liquid on-site self-energy because 
of the virtual fluctuations into the three $\Psi^{(a)}$ (with the same diagram
as Fig.3, except that site $i$ is the same as site $j$). In this regime
then, our model has 2 distinct quasiparticle relaxation times:
\bea
\Gamma_{\Phi} = - Im \Sigma_{\Phi}(\omega) \propto \omega^2 + \pi^2 T^2  \\
\Gamma_{\Psi} = - Im \Sigma_{\Psi}(\omega) \propto max [ \mid \omega \mid, T ]
\eea
 This suggests an intriguing
link to the two relaxation time phenomenology observed in the electrical
and Hall conductivities of the cuprates\cite{2tau}, where electrical conductivity
is dominated by the slower relaxation rate, while the Hall conductivity is
proportional to the {\em product} of the two relaxation times.
Unfortunately, in our model, the conserved current
$j^a$ does not include the $\Phi$ fermion, thus 
transport quantities constructed from this $O(3)$ current do not 
reveal the slow, quadratic relaxation. But another conserved quantity
in our model is the total energy. Following the same strategy as for the
$O(3)$ current, it can be shown that the conserved thermal current is
just a sum of currents due to each of the $\Phi$ and $\Psi^{(a)}$. 
Now the propagators are diagonal in the 
$\Phi$ or $\Psi$ operators, thus the thermal conductivity
proportional to the thermal current-current correlator will just be a
sum of the relaxation times:
$\kappa /T \propto \frac{3 t^2}{\Gamma_{\Psi}} + \frac{t_0^2}{\Gamma_{\Phi}}$.
Also, any
mixed correlators $\langle j^a Q^0 \rangle$ will be identically zero,
where $Q^0$ is the thermal current due to $\Phi$.
In summary, it is unfortunately impossible to see the two
relaxation times entering {\em multiplicatively} in any 
transport quantities of our model: the various conductivities derived from
the $O(3)$ current will only depend on $\Psi^{(a)}$, whereas the thermal
conductivity will be dominated by the largest relaxation time: that
of the $\Phi$ fermion.

All of our previous considerations assume that the system does not
develop into a broken symmetry state at low temperatures. In fact,
as our Hamiltonian is a generalisation of the Hubbard model at
half-filling, it is perhaps not surprising that it displays
similar magnetic or charge ordering instabilities, due to Fermi 
surface nesting. The main qualitative difference is the
presence of a large marginal Fermi Liquid regime in
the $T-U$ phase diagram. 

The plan of the paper is as follows: in Section III, we set out
the formalism of Dynamical Mean Field Theory for solving the lattice
model in the weak coupling regime. Section IV presents the results and
Section V discusses the lattice coherence scale, and the
relationship to the two-channel Kondo
lattice. We also discuss the low temperature phase of the lattice when
the Fermi surface has strong nesting instability. Finally, we touch on
the difficulties this model faces in modelling the cuprates.

\section{METHOD}

We study the Hamiltonian:
\bea
H  & = & i \tilde{t}_0 \sum_{j,c} \Phi_{j+c} \Phi_j
+ i \tilde{t} \sum_{j,c} \sum_{a=1}^3  \Psi^{(a)}_{j+c}
\Psi^{(a)}_{j} \nonumber \\ 
 & & -U \sum_j \Phi_j \Psi^{(1)}_j \Psi^{(2)}_j \Psi^{(3)}_j .
\eea where each fermion is a canonical Majorana fermion: 
$\{\Psi^{(a)}_i ,\Psi^{(b)}_j \} =
\delta_{ab} \delta_{ij} $, $\{\Phi_i ,\Phi_j\} = \delta_{ij}$, and 
$\{\Psi^{(a)}_i , \Phi_j\}= 0$.

To gain some insight into the properties of the model
in the weak coupling limit, we use
dynamical mean field theory (DMFT)\cite{Kotliar}, a method
suited to systems where the dominant interaction is on-site and 
spatial fluctuations are unimportant, and when the on-site temporal
fluctuations at all energy scales are to be taken into account.    The
$d \rightarrow \infty $ limit (where DMFT is exact),  requires the
usual scaling $\tilde{t}_a =  t_a /\surd d $ , with $t_a, U \sim
O(d^0)$. For a lattice of localised bound states $\Phi_j$, $\tilde{t}_0 = 0$.

The crucial simplification in DMFT 
is that the self energies are $k$-independent as any intersite
diagram (such as in Fig.3) is at most of order $1/\surd d$  relative to
on-site diagrams. Now the diagram of Fig.3 is precisely that which causes
the $\Phi$ fermion to propagate: omitting it means that $t_0$ remains at
zero, and $\Phi_i$ stays localised, strictly in
infinite spatial dimension. In finite $d$, we will need to
incorporate its effect, since propagating $\Phi_i$ fields will lead to the 
destruction of the marginal scattering mechanism. A  rigorous $1/d$ expansion
appears to be formidable. Nevertheless, the essential effects of
finite dimensions may be included by introducing a finite value for 
$\tilde{t}_0$, and treat it as a fixed parameter of the model. Defining 
$\tilde{t}_0 = t_0/\surd d$,
we estimate $t_0$  by calculating the diagram (Fig.3) in finite $d$ at $T=0$.
In Appendix A, we show that  $t_0/t = 
c (U/t)^2 /d$, with $t_a = t$  for $a=1,2,3$, and $c$ is a small constant.
Finally, define a zeroth component $\Psi^{(0)}_j \equiv \Phi_j$ to get:
\begin{eqnarray}
H & = & i \sum_{j,c} \sum_{a=0}^3 \frac{t_a}{\surd d} \Psi^{(a)}_{j+c} \Psi^{(a)}_{j} - U \sum_j \Psi^{(0)}_j \Psi^{(1)}_j \Psi^{(2)}_j \Psi^{(3)}_j ,\cr
t_a & = & \left\{ \begin{array}{r@{\quad:\quad}l}
t & a=1,2,3 \\
t_0 & a=0 .  \end{array} \right. 
\end{eqnarray} We shall use both $\Phi_j$ and $\Psi^{(0)}_j$ interchangeably.

Following standard procedures of DMFT\cite{Kotliar}, we map the lattice
problem to an effective {\em single-site} problem with the effective action:
\begin{eqnarray}
S_{eff} & = & \int_0^{\beta} d\tau d\tau' \{ \sum_{a=0}^3 \Psi^{(a)}(\tau)
{\cal G}^{-1}_{a} (\tau - \tau' ) \Psi^{(a)}(\tau') \nonumber \\
	  &  & - U \int_0^{\beta} d\tau
(\Psi^{(0)} \Psi^{(1)} \Psi^{(2)} \Psi^{(3)})(\tau),
\end{eqnarray} where ${\cal G}_a(\tau)$ is the dynamical mean field
at the single site which includes time dependent influence of the rest
of the lattice, and is {\em not} the original lattice non-interacting
local Green's function. It plays the role analogous to the Weiss mean
field for conventional mean field theory, and has to be determined
self-consistently: see below.

The effective single-site problem dressed Green's
function $G$ is related to $\cal G$ via:
\begin{equation}
\Sigma_a (i\omega_n) = {\cal G}^{-1}_a (i\omega_n) - G^{-1}_a
(i\omega_n) , \quad a=0,1,2,3 ,
\end{equation} where the self-energies $\Sigma_a$ are calculated from
$S_{eff}$. To lowest order in $U$, they are given by the following
diagrams in Fig.4. 

\begin{figure}[here]
\unitlength1.0cm
\vskip 0.3 truein
\begin{center}
\begin{picture}(7,4.5)
\epsfxsize=6.0cm
\epsfbox{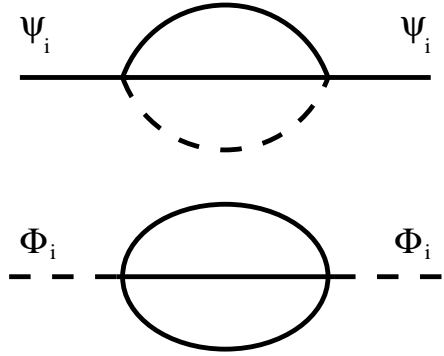}
\end{picture}
\vskip 0.3 truein
\caption{On-site self-energies. }
\end{center}
\end{figure}

The effective single site dressed Green's function must
then be related back to the $k-$average of the original
lattice dressed Green's function $G_a(k,i\omega)$, via 
the mean field self-consistency equations\cite{Kotliar}:
\begin{displaymath}
G_a(i\omega_n)  =  \int \frac{d^dk}{(2\pi)^d} G_a(k,i\omega_n)  ,
\end{displaymath} where since in $d \rightarrow \infty$ limit,
self-energies have no $k-$dependence, $G_a(k,i\omega_n) = 1/(i\omega_n
- \epsilon_a(k) - \Sigma_a(i\omega_n)) $. Doing the integration then
gives\cite{Kotliar}:
\begin{equation}
{\cal G}^{-1}_{a}(i\omega_n)   =  i\omega_n + i t_a sgn(\omega_n) , \quad
a=0,1,2,3  
\end{equation}
for the Lorentzian density of state
(DOS): $D_a(\epsilon) = t_a/( \pi (\epsilon^2 + t_a^2) )$
(corresponding to infinite range hopping); or alternatively:
\begin{equation}
{\cal G}^{-1}_{a}(i\omega_n)  = i\omega_n
- t_a^2 G_a (i\omega_n) , \quad a=0,1,2,3
\end{equation}for the semi-circular DOS
$D_a(\epsilon) = \frac{1}{\pi t_a} \sqrt{1-(\epsilon/2t_a)^2} $
(corresponding to nearest-neighbour hopping on a Bethe
lattice). Equations 17-19 and 20 or 21, and Fig.4 together define our DMFT.
The Lorentzian DOS is tractable analytically as the self-consistency
equations are decoupled from $G$; however this also means that the
effect of the lattice enters rather trivially as just a
renormalisation of the bandwidth. To check these results
we also use the semi-circular DOS where the
self-consistency equations are solved
computationally  using iterated perturbation theory\cite{Georges}.

One quantity of particular interest in the context of the marginal Fermi 
liquid is the optical conductivity.
As mentioned in Section II, the Hamiltonian of Eqn.17 does not conserve particle
number unless $t_0 = t$, thus the ordinary electrical current
proportional to the particle current is not useful. We can however generalise the 
concept of the optical conductivity to this model, using the fact that the 
total isospin\cite{Mattis} $S^a = \sum_i \frac{-1}{2} 
\Psi^{(\alpha)}_i T^a_{\alpha \beta} \Psi^{(\beta)}_i $   is conserved,
where $T^a_{\alpha \beta} = i \epsilon_{a \alpha \beta}$ are the $O(3)$ generators.
Combining the continuity equation:
$i \partial_{\tau} S_i^b + \sum_{\hat a} ( j^b_{i+{\hat a}}- j^b_{i}) = 0 $
(${\hat a}$ are the unit lattice vectors), and the
equation of motion $\partial_{\tau} S_i^b = [H , S_i^b] $ leads to
 the conserved current:
\begin{equation}
 j^b_{i+{\hat x}} = i t \epsilon_{b \alpha \beta} 
\Psi^{(\alpha)}_{i+{\hat x}} \Psi^{(\beta)}_i  .
\end{equation} This is the Noether current associated with the
$O(3)$ symmetry of our Hamiltonian. We can then introduce a vector potential
field $\vec{A}^{(a)} (\vec{x}) = (A^{(a)}_1 A^{(a)}_2 \cdots A^{(a)}_d)$ in $d$ 
dimensional space, coupled to the electrons as follows:
\bea
H_A = i t \sum_{i,c} \Psi^{(\alpha)}_{i+c} \exp \left\{ i 
\int d\vec{l} \cdot \vec{A}^{(a)} T^a \right\}_{\alpha \beta}  
\Psi^{(\beta)}_i,
\eea 
where the line integral goes from site $i$ to $i+c$,
and a summation is implied over dummy indices.  Since there is isotropy for 
$a=1,2,3$, we need only study the response
to the the $\vec{A}^{(1)}$ component
$j^{(1)}_x (\omega) = \sum_y \sigma_{x y}(\omega) A^{(1)}_y (\omega) $, which can be described
by a Kubo formula\cite{Mahan}:
\begin{eqnarray}
\sigma_{xx}(i\omega_n) &=& \left.  \frac{1}{-\omega_n} \Pi(\vec q, i\omega_n)
\right|_{\vec q=0} ,
\nonumber \\
\Pi(\vec q, i\omega_n) &=& -\int_0^{\beta} d\tau e^{i\omega_n
\tau} \langle T_{\tau} j^{\dagger}_x(\vec q, \tau) j_x(\vec q,0) \rangle  .\nonumber
\end{eqnarray} 
In the $d \rightarrow \infty$ limit, the absence of vertex 
corrections to the conductivity bubble\cite{Kotliar} permits us to write:
\begin{displaymath}
\sigma(\nu_m) = \sum_{\vec k} T \sum_{\omega_n} (v_{k_x})^2
G_a(k, i\omega_n + i\nu_m) G_a(k, i\omega_n). 
\end{displaymath} As usual, at temperatures much lower than the
bandwidth, doing the Matsubara sum leads to a function peaking largely
near $k_F$, and we replace $\sum_{\vec k} (v_{k_x})^2$ by $n/m \int
d\epsilon$. Doing the energy integral and analytically continuing to
real frequencies:
\begin{eqnarray}
\sigma(\nu + i \delta )  =  \frac{n}{m} \int^{+\infty}_{-\infty}
d\omega \left[ \frac{f(\omega_-) -  f(\omega_+)}{-i\nu} \right] \nonumber \\
   \times  \left[ \nu - (\Sigma_R(\omega_+)-\Sigma_R(\omega_-) ) + i
(\Gamma (\omega_+) + \Gamma (\omega_-)) \right]^{-1} 
\end{eqnarray} where $\Sigma (\omega \pm i 0^+) = \Sigma_R (\omega)
\mp i \Gamma (\omega)$ and $\omega_{\pm} = \omega \pm \nu /2$.

\section{RESULTS}

In the $d \rightarrow \infty$ limit (ie. $t_0 = 0$), ${\cal G}_0$
is the same as the single impurity model, by Eqn 19 or 20, so the bound-states
described by $\Psi^{(0)}_i$ are localised, with a 
self-energy that has a Fermi liquid form:
\bea
\Sigma_{\Phi} (\omega^+ ) =  - (U N_0)^2 \left\{ \frac{4 \omega}{\pi}
+ i \frac{\pi N_0}{2} ( \omega^2 + (\pi T)^2 ) \right\},
\eea
where $\omega^+=\omega+ i 0^+$ and 
$N_0$ is the DOS at the Fermi surface.
We give a brief derivation of this result in Appendix B.
For $\Psi^{(a)}_i, a=1,2,3$, the mean field
${\cal G}_a$ of the effective single site problem  is of the same form as
the single impurity model bare Green's function, using the Lorentzian
DOS for the $d \rightarrow \infty$ lattice. This is due to the
DOS in the effective problem being smooth at Fermi energy, and for
$T \ll t$, equaling a constant $N_0$, just as in the single impurity model\cite{2CCK} We obtain the marginal self-energy:
\begin{eqnarray}
\Sigma_{\Psi} (\omega^+)  = 
- (U N_0)^2 \omega \left[ \ln
\frac{\Lambda}{T} - \Psi \left(1 - \frac{i\omega}{ 2 \pi T}\right)
 + {\pi T\over i\omega} \right],  
\end{eqnarray}
where $\Lambda$ is a cut-off proportional to $t$, $\Psi$
is the Digamma function, $\gamma \sim 0.6$ is the Euler constant (Appendix B).
This has the following limiting behavior
\begin{equation}
\Sigma_{\Psi} (\omega^+)  = 
  (U N_0)^2 \times \left\{ \begin{array}{r@{\qquad}l}
(-\omega \ln \frac{2 \pi \Lambda}{|\omega|} - 
		i \frac{\pi}{2} |\omega|), &  (\omega \gg T) \\
                  & \\
		(-\omega \ln \frac{\Lambda e^{\gamma} }{T} - 
		i \pi T), & (\omega \ll T)  \end{array} \right.
\nonumber
\end{equation}

At finite $d$, the  lattice coherence energy scale
$t_0$ generated from the diagram of Fig.3 becomes
finite, with $t_0 \sim U^2/t$.  
Marginal Fermi liquid behaviour will now
persist so long as $t_0 < T \ll t$ . At lower temperatures
$T \ll t_0 < t$ the three-body
bound-states begin to propagate coherently, causing a cross-over
to Fermi liquid behavior. 
This is borne out by analytical calculations; here we
illustrate with computational results (using the semi-circular DOS) in
Fig.5 showing the effective quasiparticle scattering rate $\Gamma
(\omega) = \omega Re\sigma (\omega) / Im \sigma (\omega)  
\sim -Im \Sigma_a (\omega^+)
$, where $\sigma (\omega) $ is the optical conductivity defined in
Methods. In Fig.6 we plot the the resistivity $\rho (T)$ showing the
large linear $T$ regime at weak coupling, and the inset shows the
crossover to the  $T^2$ Fermi Liquid regime.  

\begin{figure}
\unitlength1.0cm
\begin{center}
\begin{picture}(9,6)
\epsfxsize=9.0cm
\epsfbox{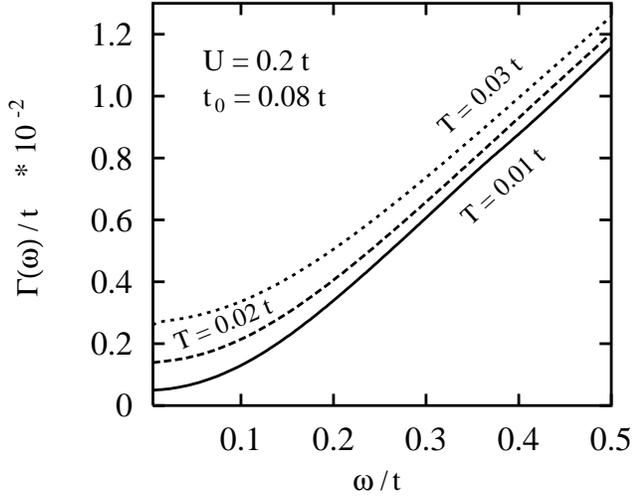}
\end{picture}
\vskip 0.4 truein
\caption{Plot of quasi-particle scattering rate $\Gamma
(\omega) = \omega Re\sigma/Im\sigma $. }
\end{center}
\end{figure}

\begin{figure}
\unitlength1.0cm
\begin{center}
\begin{picture}(9,6)
\epsfxsize=9.0cm
\epsfbox{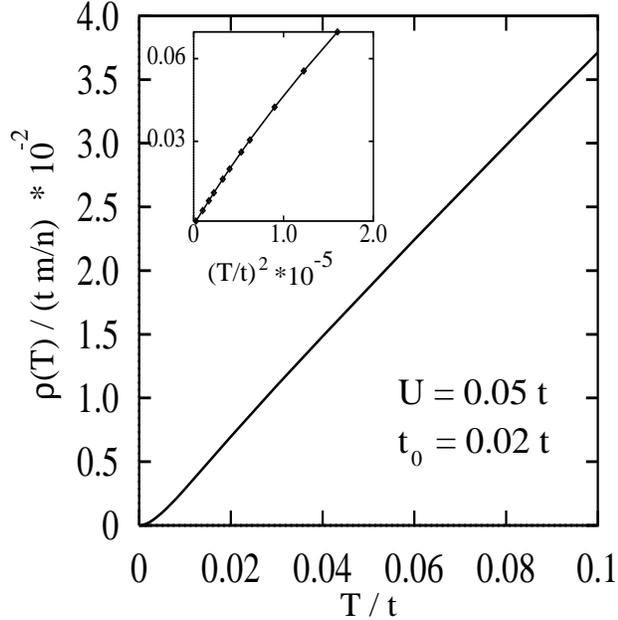}
\end{picture}
\vskip 0.4 truein
\caption{Plot of resistivity vs.$T$. Inset shows the very low 
temperature crossover to $T^2$ behaviour: the y-axis is in 
the same units as the big plot.}
\end{center}
\end{figure}

\section{DISCUSSION and CONCLUSION}


In this paper we have shown how a lattice of three-body bound-states
induces marginal Fermi liquid behavior above a lattice
coherence temperature $t_o$ where a 
Fermi liquid forms. Since $t_0 \sim U^2 / t $, a substantial window in
temperature for marginal Fermi Liquid behaviour exists only for small $U$. 
The emergence of this lattice coherence energy is expected to be quite general: 
whenever a
localised mode is allowed to interact with conduction
electrons, it will be difficult to prevent hybridizations between 
fields at different sites. These effects however, are missed in 
a strict $d \rightarrow \infty$ calculation. 

This brings us onto the question of the relationship between our
lattice model and the two-channel Kondo lattice model.  The
single-impurity version of the Hamiltonian of Eqn.11 (with $U < 0$)
was originally derived\cite{2CCK} in the context of the
single-impurity two-channel Kondo model\cite{Nozieres}. That
derivation took advantage of spin-charge separation\cite{Emery} to
throw away uncoupled (charge) degrees of freedom, and it has been
shown via bosonization to be exactly equivalent to the original
model\cite{Schofield}.  Unfortunately there is no such relation
between the two-channel Kondo lattice and the Majorana lattice
considered here.

In the above calculations, we have assumed that the marginal Fermi
Liquid state is unstable only to the Fermi Liquid state at low $T$. In
fact, just as in the Hubbard model , a mean field calculation
indicates that for a Fermi surface with a strong nesting instability
(for example, nearest neighbour hopping in a hypercubic lattice),
there is a phase transition to antiferromagnetic order (for
$U>0$). The order parameter is a vector that  reflects the
$O(3)$ symmetry of the model:
\begin{eqnarray}
V^a(\vec x_j) & = & e^{i Q \cdot x_j} \langle c^{\dagger}_{\alpha}
\sigma^{a}_{\alpha \beta} c_{\beta} \rangle \nonumber \\
  & = & i e^{i Q \cdot x_j} \left\langle \Psi^{(0)} \Psi^{(a)} -
\frac{1}{2} \left( {\vec \Psi} \times {\vec \Psi} \right)^a \right\rangle ,
\end{eqnarray} where $Q=(\pi,\ldots ,\pi)$ is the nesting vector.
From the divergence
in the susceptibility, we find that
at weak coupling, $T_c < t_0$, except when $t_0 = 0$: 
\begin{equation}
\frac{T_c}{\Lambda} = \exp \left[ \frac{\ln t_0/t}{4} - \sqrt{
\frac{(\ln t_0/t)^2 }{4} + \frac{1+t_0/t}{2 (U N_0)^2}} \right]  , 
\end{equation}where $\Lambda$ is a cut-off ($\Lambda < t$).  (Note
that this reduces to the Hubbard model value when $t_0=t$.) At $t_0 =
0$, $T_c$ is identical to the Hubbard case. For  $0<t_0<t$,
$T_c$ is enhanced relative to the Hubbard case.  Hence a region of
Fermi liquid phase separates the low temperature antiferromagnetic
phase from the marginal Fermi Liquid regime: Fig.7. 

\begin{figure}
\unitlength1.0cm
\begin{center}
\begin{picture}(9,9)
\epsfxsize=9.0cm
\epsfysize=9.0cm
\epsfbox{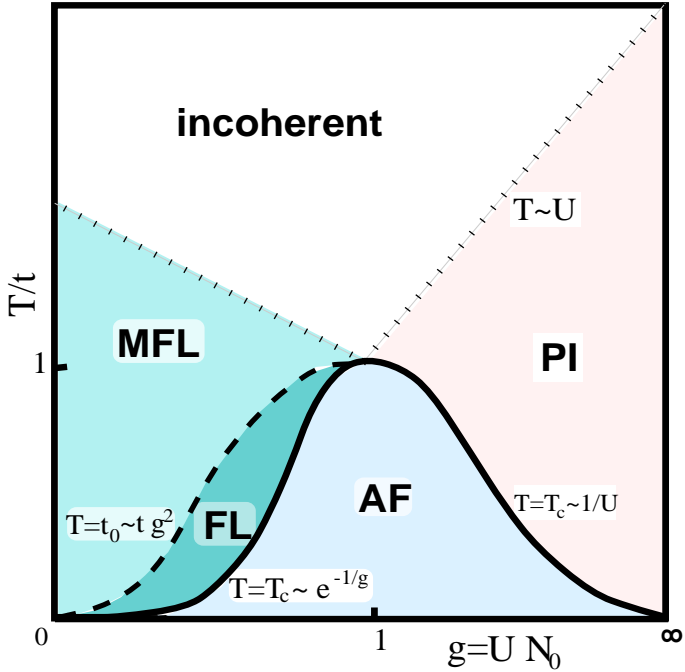}
\end{picture}
\vskip 0.4 truein
\caption{Schematic phase diagram of the Majorana lattice model. The
dimensionless coupling is $g \equiv U N_0$ where $N_0 \propto 1/t$ is the density of states
at the Fermi surface. $t_0 = c t g^2 /d$. MFL is the marginal Fermi
liquid phase, FL is the Fermi liquid phase. AF is the antiferromagnetic phase,
PI is the paramagnetic insulating phase. At weak coupling, $T_{c}$ goes as
$\exp(-1/g)$, while at strong coupling, it goes as $t^2/U$. We have no
computation for strong coupling for $T>T_{c}$, thus we do not know the
continuation of the $T=t_0$ line. However,  we might expect that when
$U$ is of order $t$, $T_{c}$ and $t_0$ will also be of order 
$t$.
One scenario is then the  $T=t_0$ line meets and ends at the $T=T_{c}$ line
around $t_0$. Also, at strong coupling, the paramagnetic insulator crosses
over to a metal at $T \sim U$ (in analogy to the Hubbard model), and in the 
weak coupling regime, there will be a similar high temperature cross-over 
from marginal Fermi Liquid to an incoherent metallic phase: these cross-overs are indicated 
with a dotted line. }
\end{center}
\end{figure}

There are further similarities to the Hubbard model at half-filling.
Both of the $SO(4)$ and $O(3)$ models are invariant under $U \rightarrow
-U$ and $\Psi^{(0)} \rightarrow -\Psi^{(0)}$. The latter map
corresponds to a particle-hole transformation for the down spin only:
$c_{\downarrow} \leftrightarrow c^+_{\downarrow}$. This implies that
in going from the positive $U$ model  to the negative $U$ model, magnetic
ordering turns into charge ordering\cite{Fradkin} . It can also be
shown that neither model mixes charge and magnetic ordering. Further,
both models reduce to the Heisenberg antiferromagnet as $U \rightarrow
\infty$. Thus our model has very similar properties to the half-filled
Hubbard model, except for the marginal Fermi Liquid phase at weak coupling.

What insight does our model bring towards the understanding of the marginal
Fermi liquid behaviour in the cuprates? While our model does provide a
simple lattice realisation of a marginal Fermi liquid, it
unfortunately suffers from a number of defects:
\begin{itemize}
\item It has a wide window of marginal Fermi liquid behaviour
only for small coupling. The cuprates are believed to be in the strong
coupling regime\cite{largeU}, but at strong coupling, our system has
charge or magnetic instabilities. Related
to this is the fact that in the cuprates, the inelastic scattering
rate $\Gamma = \Gamma_0 {\rm max}[\omega, T]$ has $\Gamma_0 /t$ a
constant of order one, whereas our model has $\Gamma_0 /t$
proportional to the coupling squared. 
\item  The model needs to be at half filling: upon doping, a chemical
potential term $\mu (\Psi^{(0)} \Psi^{(3)} + \Psi^{(1)} \Psi^{(2)})$
leads to a width $\Delta \propto \mu^2$ for $\Phi$, with Fermi liquid
properties when $T<\Delta$. This seems to require fine tuning,
contradicting the rather robust linear-T resistivity observed even in
underdoped systems (above the ``spin gap'' scale). 
\item Despite the presence of two relaxation rates in the system, transport
quantities will not involve a multiplicative combination of the
$\Phi$ and $\Psi$ relaxation rates, as is postulated in the two-relaxation-times
phenomenology for the cuprates. (See Section II.)
\end{itemize}

In conclusion we have demonstrated the persistence of marginal Fermi Liquid behaviour at weak
coupling in a toy model of a marginal Fermi Liquid in an infinite dimensional lattice. 
For finite $d$, the lattice coherence energy cuts off marginal Fermi Liquid behaviour
and the system   reverts to a Fermi Liquid at low temperatures. 
Since this cut-off grows with the coupling, there will be no marginal Fermi Liquid
regime at strong coupling. It remains to be seen if a strong coupling
marginal Fermi Liquid exists in  any finite dimensions.
\vskip 0.1truein
{\em Acknowledgement}
We acknowledge useful discussions with Andrew Schofield, Gunnar P\'alsson
and Revaz Ramazashvili. This work was supported by NSF grant DMR-96-14999. 
Part of the work was done while the authors were in the Non-Fermi Liquid
Workshop at the Institute of Theoretical
Physics (ITP), UCSB, funded under NSF grant PHYS94-07194 and DMR-92-23217.
 We thank the staff at ITP for their hospitality.

\section{APPENDIX A: Calculation of the effective bandwidth $\lowercase{t}_0$}

We want to estimate in finite dimensions the effective kinetic energy $i \tilde{t}_0 \sum_{i,c} 
\Psi^{(0)}_{i+c} \Psi^{(0)}_i$ from the zero-frequency part of 
the $\Psi^{(0)}$ self-energy, as depicted in Fig.3, to lowest order in the
coupling:
\bea
i \tilde{t}_0 =  U^2 \int_0^\beta d\tau [G_{\hat x}(\tau)]^3
\eea
where $G_{\hat x}(\tau)$ is the bare propagator of $\Psi^{(a)} (a=1,2,3)$, for nearest
neighbour sites, taken here to be in the ${\hat x}$ direction. In $k, \omega$ space,
$G({\vec k}, i\omega_n) = [i\omega_n - \epsilon_{\vec k}]^{-1}$. For simplicity,
take $\epsilon_{\vec k} = -2 \tilde{t} \sum_{i=1}^d \sin (k_i)$, appropriate for the 
hypercubic nearest neighbour dispersion. (We expect that as $T \rightarrow 0$, the
exact shape for the band does not matter.) Let $\tilde{t} = t/\surd d$ and 
$\tilde{t}_0 = t_0/\surd d$, as required 
for a proper scaling of the kinetic energy term in large $d$ limit. Then $G_{\hat x}(\tau)$
will be of order $1/\surd d$, and as mentioned in 
the Methods section, $\tilde{t}_0 \sim O(d^{-3/2})$, ie., $1/d$ down on
the dispersion for the $a=1,2,3$ components. Doing the standard Matsubara sum leads to:
\bea 
G_{\hat x}(\tau) = \int \frac{d^d k}{(2 \pi)^d}  
f(-\epsilon_{\vec k})  \exp(i {\vec k} \cdot {\hat x} -\epsilon_{\vec k} \tau) .
\eea
In the zero temperature limit, the Fermi function become $f(-\epsilon_{\vec k}) 
\rightarrow \theta (\epsilon_{k_x} + \epsilon^{\prime}_{\vec k} )$, where we have split up 
the dispersion into the $k_x$ part and the other $d-1$ part. Turning the $d-1$ 
dimensional $k-$integral into an energy integral gives:
\bea
G_{\hat x}(\tau) = \int \frac{dk_x}{(2 \pi)} e^{i k_x - \epsilon_{k_x} \tau}
\int_{-\epsilon_{k_x}}^{\infty} d\epsilon N(\epsilon) e^{-\epsilon \tau}  ,
\eea where $N(\epsilon)$ is the density of states in $d-1$ dimensions. 

To make further progress, a flat density of states is used: $N(\epsilon) = 
1/(4 t)$ for $|\epsilon| < 2 t$ and zero otherwise. Thus,
defining the dimensionless time variable $s = 2 \tau t$,
\bea
G_{\hat x}(s) = - \frac{e^{-s}}{2 s} J_1 (i \frac{s}{\surd d}),
\eea
where $J_{1}(z) = \int_{-\pi}^{\pi} \frac{dx}{2 \pi} e^{-i z \sin(x) + i  x} $ is the 
Bessel function of the first kind of the first order. Thanks to the factor of
$1/\surd d$ inside its argument, the Bessel function asymptotically always grows 
more slowly than the decay due to the  $ e^{-s} $ factor. (For $|arg(z)| < \pi$ and
$|z| \rightarrow \infty$, $J_1(z) \rightarrow (\frac{2}{\pi z})^{1/2} \cos(z - 3 \pi /4)$.)
Thus, contributions to the
integral in Eqn.30 are dominated by the regime when the Bessel function is 
at most of order one, allowing us to 
approximate: $i J_1 (i x) = -x/2 + O(x^3)$ in Eqn.33, leading finally to:
\bea
\tilde{t}_0 \simeq \frac{U^2}{2 t} \int^{\infty}_0 ds
\left[ \frac{e^{-s}}{2 s} \frac{s}{2 \surd d} \right]^3 \nonumber \\
= \frac{U^2}{2^7 \cdot 3 \quad t \quad d^{3/2}} ,
\eea with $\tilde{t}_0$ of order $d^{-3/2}$ as claimed.

\section{APPENDIX B: calculation of the marginal self-energy}

To order $U^2$, the on-site self-energy for $\Psi^{(a)}_i$, $a=1,2,3$ is:
\bea
\Sigma_{\Psi} (\tau) = U^2 {\cal G}_0 (\tau)  {\cal G}_a (\tau) {\cal G}_a (\tau) .
\eea
${\cal G}_0 (\tau) = sgn(\tau) /2 $ is identical to the single impurity $\Psi^{(0)}$ 
propagator\cite{2CCK}, since in the strict $d \rightarrow \infty$ limit, $t_0 = 0$. 
For $\Psi^{(a)}_i$, 
\bea
{\cal G}_a (\tau) = T \sum_{\omega_n} 
\frac{e^{-i\omega_n \tau}}{i\omega_n + i t_a sgn(\omega_n)}
\eea where we have used Eqn.19 for ${\cal G}_a (i\omega_n)$ for the Lorentzian DOS.
As usual, turn the Matusbara sum into a contour integral and deform the contour onto
the branch cut at the real axis to get:
\bea
{\cal G}_a (\tau) = - Im \int_{-\infty}^{\infty} \frac{d\omega}{\pi} (1-f(\omega)) 
\frac{e^{-\omega \tau}}{\omega + i t_a} .
\eea ($f(\omega)$ is the Fermi function.) As we are interested in $T\ll t_a$, the
integrand is dominated by small $\omega$, and we approximate the 
denominator $\omega +i t_a \approx
i t_a$. Now the $d-$dimensional Lorentzian DOS at the Fermi surface is $N_0 = 
1/(\pi t_a)$, and thus ${\cal G}_a (\tau)$ is identical to that for the single
impurity  calculation\cite{2CCK}:
\bea
{\cal G}_a (\tau) = \frac{N_0 \pi T}{\sin(\tau \pi T)} .
\eea Note that this expression is accurate for $0 \ll \tau \ll \beta$. Going to Matsubara
frequencies $\omega_n = (2n+1) \pi T$:
\bea
\Sigma_{\Psi} (i \omega_n) = U^2 \int_0^{\beta} d\tau \frac{sgn(\tau)}{2} \left(
\frac{N_0 \pi T}{\sin(\tau \pi T)} \right)^2 e^{-i \omega_n \tau} \nonumber \\
= -i \frac{\pi T}{2} (U N_0)^2 I_n(\epsilon) 
\eea where 
\bea
I_n(\epsilon) = \int_{\epsilon}^{\pi -\epsilon} dx \frac{\sin (2 n + 1)}{(\sin (x))^2} .
\eea We have put in a cut off $\epsilon = \pi T/(2 t_a) \ll 1$. Integrating
by parts twice, and using the tabulated integral\cite{GR}: 
$\int_0^{\pi} dx  \ln \sin (x) \sin (2 n + 1)x = -
\frac{2}{2n+1} (\frac{1}{2n+1} + \ln 2 + \gamma + \Psi (n+1/2) )$, we get:
\bea
I_n (\epsilon) = \frac{2 \omega_n}{\pi T} \left( \ln \left(\frac{\Lambda}{T}\right) - 
\Psi \left(\frac{\omega_n}{2 \pi T} \right) \right) -2 .
\eea $\Psi(x)$ is the Digamma function, $\Lambda = t_a e^{1-\gamma}/(\pi T)$ and
$\gamma \sim 0.6$ is the Euler constant. (We have expanded in $\epsilon$ and kept only
the terms up to $\epsilon^0$.)
Putting this all together and performing the
analytic continuation $i\omega_n \rightarrow \omega + i 0^+$ gives Eqn. 26.

The on-site self-energy for $\Phi$ is:
\bea
\Sigma_{\Phi} (\tau) = U^2 {\cal G}_a (\tau)  {\cal G}_a (\tau) {\cal G}_a (\tau) .
\eea Fourier transforming:
\bea
\Sigma_{\Phi} (i \omega_n) = i N_0 (N_0 U \pi T)^2 K_n , \\
K_n = \int_{\epsilon}^{\pi -\epsilon} dx \frac{ \sin (2 n +1) x}{(\sin (x))^3}. \nonumber
\eea Again, we have a cut-off $\epsilon = \pi T/(2 t_a) \ll 1$. Integrating by parts and
expanding in $\epsilon$, $K_n = 2 (2 n +1)/\epsilon - 2 n (n +1) \pi$. This then leads to
Eqn. 25 after analytic  continuation.

\end{document}